\documentclass[a4paper,10pt]{article}
\usepackage[utf8]{inputenc}
\usepackage{amssymb,amsmath,epsfig}
\title{Fidelity Susceptibility as Holographic PV-Criticality}
\author{Davood Momeni$^{a}$, Mir Faizal $^{b,c}$,\\Kairat Myrzakulov $^{a}$ , Ratbay Myrzakulov$^{a}$\\\\
$^{a}$ Eurasian International Center for Theoretical Physics \\ and
Department of General Theoretical Physics,\\ Eurasian National
University, Astana 010008, Kazakhstan.\\\\
$^{b}$ Department of Physics and Astronomy, \\University of Lethbridge,\\ Lethbridge, Alberta T1K 3M4, Canada.
\\\\
$^{c}$ Irving K. Barber School of Arts and Sciences, 
\\ University of British
Columbia - Okanagan,  
3333 University Way,\\  Kelowna,   British Columbia, V1V 1V7, Canada.}
\date{}
\begin{document}

\maketitle

 \begin{abstract}
It is well known that entropy can be  used to holographically establish a connection between geometry,  thermodynamics and 
information theory. In this paper, we will use complexity to holographically   establish a connection between geometry, 
thermodynamics and information theory. Thus, we will analyse the relation between holographic complexity, 
fidelity susceptibility, and thermodynamics in extended phase space. 
We will demonstrate that  fidelity susceptibility (which is the informational complexity  dual to a maximum 
volume in AdS) can be related to the thermodynamical volume (which is conjugate to the cosmological constant in the extended 
thermodynamic phase space).  
Thus,  this letter establishes a relation between geometry, thermodynamics, and information theory, 
using complexity. 
\end{abstract}

Studies done on various different branches of physics seem to indicate that the physical laws 
are informational theoretical processes, as they can be represented by the ability of an observer to process 
  relevant information \cite{info, info2}.  However, in such a process it is important to know the amount 
  of information that can be processes, and hence, the concept of loss of information in a informational 
  theoretical process become physically important. This  loss of information in a process is quantified using 
  the concept of   entropy.  It may be noted that even the structure of spacetime can be viewed as an emergent 
  structure which occurs due to a certain scaling behavior of entropy in the 
  Jacobson formalism  
 \cite{z12j, jz12}. In fact, in this formalism general relativity is obtained by using this scaling behavior 
 of maximum entropy i.e., the maximum entropy of a region of space scales with its area. This scaling behavior 
 of maximum entropy is motivated by the physics of black holes, and it in turn motivates the holographic principle
 \cite{1, 2}. The holographic principle equates the   number of degrees of freedom in a region of space
  to the number of degrees 
of freedom on the boundary surrounding that region of space. The   AdS/CFT correspondence is  of the most
important realizations of the holographic principle \cite{M:1997}. It relates the supergravity/string theory 
in AdS spacetime to the superconformal field theory on the boundary of that AdS  spacetime. 
It may be noted that AdS/CFT correspondence has been used to obtain quantify the entanglement by using the concept of 
quantum entanglement entropy, and this has in turn been used to address the black hole information paradox   \cite{4,5}.
 Thus,   for a  
   subsystem $A$ (with its complement), it is possible to define    $\gamma_{A}$ as the  $(d-1)$-minimal surface extended 
into the AdS bulk with the boundary $\partial A$, and the holographic entanglement entropy for this system can be written as 
  \cite{6, 6a}
\begin{equation} \label{HEE}
\mbox{Entropy}_{A}=\frac{\mathcal{A}(\gamma _{A})}{4G_{d+1}}
\end{equation}
where $G$  is the gravitational constant in the AdS spacetime, and $\mathcal{A} (\gamma_A)$ is the area of the minimal surface. 
This relation can be viewed as 
a connection between a geometry, thermodynamics and information theory. As it relates a geometrical quantity (minimal volume), 
to a thermodynamical quantity (entropy), and which in turn related to information theory (loss of information). 
In this paper, we will analyse this correspondence further, and establish such correspondence between volume in AdS,  the 
difficulty to process information, with a thermodynamical quantity.

In a information theoretical process, it is not only important to know how much information is retained in a system, 
but also how easy is it to obtain that information. Just    as entropy quantifies loss of information, a new quantity 
called the complexity 
quantifies  the difficulty to obtain that information.  Now as the physical laws are thought to be informational 
theoretical processes,  complexity (just like entropy) 
is a fundamental quantity.  In fact,   
 complexity   has been used to analyse  condensed matter systems \cite{c1, c2},   molecular physics \cite{comp1}, and even 
 quantum computational systems  \cite{comp2}. Recently studies done on black hole physics seem to indicate that complexity 
 might be very important in understanding the black hole information paradox, and this is because 
 the  information may not be ideally lost in a black hole, however, as 
 it would be impossible to reconstruct it from the Hawking radiation, it would be effectively lost \cite{hawk, chao, moffat}. 
It has been suggested that the complexity would be dual to a volume in the bulk AdS spacetime, 
\cite{Susskind:2014rva1,Susskind:2014rva2,Stanford:2014jda,Momeni:2016ekm}, 
\begin{equation}\label{HC}
\mbox{Complexity} = \frac{V }{8\pi R G_{d+1}},
\end{equation}
where $R$ and $V$ are the radius of the curvature and the volume in the AdS bulk. 
As there are different ways to define a volume in AdS,   different proposals for complexity  have been proposed.
For a subsystem $A$ (with its complement), it is possible to 
use the volume enclosed by the same   minimal surface which was used to calculate the holographic 
entanglement entropy,  $V = V(\gamma)$ \cite{Alishahiha:2015rta}. This quantity is usually denoted by 
$ \mathcal{C}$. 
However, we can also define the complexity using the 
  maximal volume in the AdS which
ends on the time slice at the AdS boundary, $V = V(\Sigma_{max})$ \cite{MIyaji:2015mia}. It has been demonstrated that 
the complexity calculated this way is actually the 
fidelity susceptibility   of the boundary CFT. So,   this   quantity is called as the    fidelity susceptibility
even in the bulk, and it is  denote   by 
$ \chi_F$ \cite{MIyaji:2015mia}. The   fidelity susceptibility of the boundary theory can be used for 
  analyzing   the quantum phase transitions \cite{r6,r7, r8}. Thus, like the holographic entanglement entropy, 
  this establishes a connection between geometry and information theory. As we want to distinguish between these 
  two quantities, we shall call $  \mathcal{C}$ as holographic complexity \cite{Alishahiha:2015rta}, and $ \chi_F$
  as  fidelity susceptibility \cite{MIyaji:2015mia}. 
  
  In this paper, we would like to demonstrate 
  that this connection can be extended even to thermodynamics. So,  just like holographic entanglement entropy 
  was used to establish a connection between information theory, geometry, and thermodynamics, we will demonstrate 
  that  fidelity susceptibility also establishes a connection between information theory, geometry and thermodynamics. 
  The missing part of this connection is the connection between fidelity susceptibility and thermodynamics.
To establish this connection for a concrete example, let us now consider 
 the Schwarzschild black holes in AdS backgrounds (SAdS$_4$). 
The metric is given by 
\begin{eqnarray}
ds^2=-fdt^2+\frac{dr^2}{f}+r^2(d\theta^2+\sin^2\theta d\varphi^2), \label{sch_metric}
\end{eqnarray}
where  
\begin{eqnarray}f&=& 1-\frac{2M}{r}+\frac{r^2}{\ell^2}={\frac { \left( r-r_{+} \right)  \left( {r_{+}}^{2}+rr_{+}+{l}^{2}+{r}^{2}
 \right) }{{l}^{2}r}}\nonumber \\ &=& 1+\frac{r^2}{l^2}\Big(1-\frac{2\epsilon r_{+}^3}{r^3}
\Big). 
\end{eqnarray}  
Following the standard procedure, the black hole mass $M$ is related to the temperature, through
the Wick rotation $\tau=i t$ (this requires  the resulting Euclidean geometry to be free from 
a conical singularity). Denoting the position of the horizon by the largest root of $f(r_+)=0$,
 the mass, temperature and the entropy of the black hole can be expresses as \cite{ext2}
\begin{eqnarray}
M_{BH}&=&\frac{r_{+}}{2}\left(1+\frac{r_{+}^2}{l^2}\right),\\
T_{BH}&=&
%\beta_0^{-1}=
\frac{\ell^2+3r_+^2}{4\pi\ell^2r_+},
\\
S_{BH}&=&\frac{A}{4G}
%\nn\\
%&=&\frac{\text{vol}(S^{d-2})}{4G_{d+1}}\int d\theta~ r_+(r_+\sin\theta)^{d-2}
%\nn\\
=\frac{{vol}(S^{2})}{4G}r_+^{2}.
\end{eqnarray}
Since we plan to use perturbative calculations, we assume $M$ is small and for the 
the horizon, we take   $r_+ \propto M$. We can define the horizon size in terms of temperature $T$ (we set $l=1$) as follows:
\begin{eqnarray}
&&r_{+}=l\left( \frac{2}{3}\,T\pi \,l-\frac{1}{3}\,\sqrt {4\,{T}^{2}{\pi }^{2}{l}^{2}-3}
 \right) 
\label{rplus}
\end{eqnarray}
note that temperature has a minimum locates at $T_{min}= {\sqrt{3}}/{2\pi l}$.

\begin{figure}
  \centering
  \includegraphics[width=10cm]{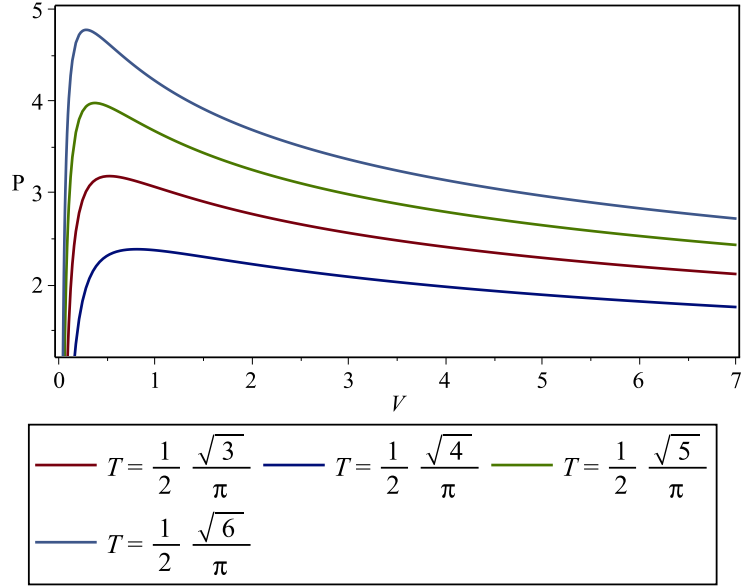}
  \caption{Graph of thermodynamical $PV$,   given by Eqs. (\ref{P},  \ref{EoS}).  }
\label{fig:thermo}
  \end{figure}

We define a perturbation parameter
$\epsilon\equiv M/{l}$, and analyse all the expression  up to first order in $\epsilon$, i.e., 
we neglect any $\mathcal{O}(\epsilon)$ contribution.
Now we can obtain  a thermodynamical  quantity,  which can be viewed as a volume in the bulk AdS spacetime. 
 It has been observed that  when a charge or 
rotation are added to a AdS black hole,  their behavior qualitatively becomes analogous  to a
Van der Waals fluid \cite{w1, w2}. This analogy between AdS black holes and a Van der Waals fluid becomes more evident 
in extended phase space, where the cosmological constant is treated as the thermodynamics pressure \cite{ext1, ext2}. 
Thus, it is important to study the  extended phase space for a system. In this paper, we will use the extended phase space, 
and relate it to the fidelity susceptibility of a system. 
Thus, in the extended phase space,   the cosmological constant  $\Lambda$ is treated as  the thermodynamic
pressure $P=- {\Lambda}/{8\pi}= {3}/{8\pi l^2}$, and   the first law of black hole 
thermodynamics is written as $\delta M=T\delta S+V\delta P$.  The  thermodynamic volume is defined as the quantity 
conjugate to $P$ 
\begin{eqnarray}
P=\Big(\frac{\partial M}{\partial V}\Big)_{S,...}, \label{P}
\end{eqnarray}
where  all other quantities like $S,...$ are fixed. Thus, it is also possible to write the  black hole 
equation of state using, $P = P (V, T )$,  and compare it to the corresponding fluid mechanical equation of state. 
It may be noted that it 
is also possible to construct a    quantity thermodynamically conjugate to pressure, and this quantity 
represents the thermodynamical volume. In this paper, we will demonstrate that this thermodynamical volume corresponds to
the fidelity susceptibility, thus establishing the connection between thermodynamics, information theory, and 
geometry.  So, now  using  metric (\ref{sch_metric}),   we observe that the 
thermodynamic 
volume can be written as  $V= {4\pi r_{+}^3}/{3}$, 
using (\ref{rplus}) and definition of $P$ we have $V=V(T,P)$, and this equation of state is given by the following expression, 
\begin{eqnarray}
&&V=\frac{1}{48}\,{\frac { \left( T\pi -\sqrt {{T}^{2}\pi -2\,P}\sqrt {\pi }
 \right) ^{3}}{{\pi }^{2}{P}^{3}}}
\label{EoS}
\end{eqnarray}
  We plot (\ref{EoS}) in  Fig.  (\ref{fig:thermo}).  
We plot $P$ as thermodynamic pressure , defined in Eq.  (\ref{P}) versus thermodynamic volume given by Eq.  (\ref{EoS}). 
This graph shows EoS of black hole for different isothermal lines when $T=\mbox{constant}$. Note that due to the EoS,
temperature is always bounded to $T\geq \frac{1}{2}\frac{\sqrt{3}}{\pi}$.  This graph  demonstrates that the pressure
initially increases with volume, and then after reaching a maximum volume it slowly decreases with the further increase in volume.
We will now compare this behavior of thermodynamic  pressure and its 
conjugate volume to the pressure and volume defined for different information theoretical complexity, and observe that this
behaviour of thermodynamic  pressure and  volume matches the behavior of the pressure and volume defined from 
the  fidelity susceptibility of the system.

\begin{figure}
  \centering
  \includegraphics[width=10cm]{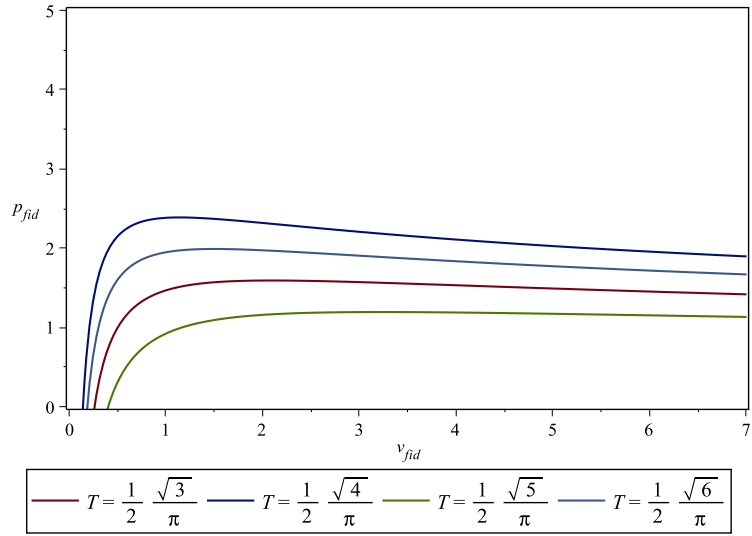}
  \caption{ Graph of  $p_{fid}v_{fid}$,  given by Eqs.  (\ref{Pfid}, \ref{Vfid}).  }
\label{fig:fid}
  \end{figure}

\begin{figure}
  \centering
  \includegraphics[width=10cm]{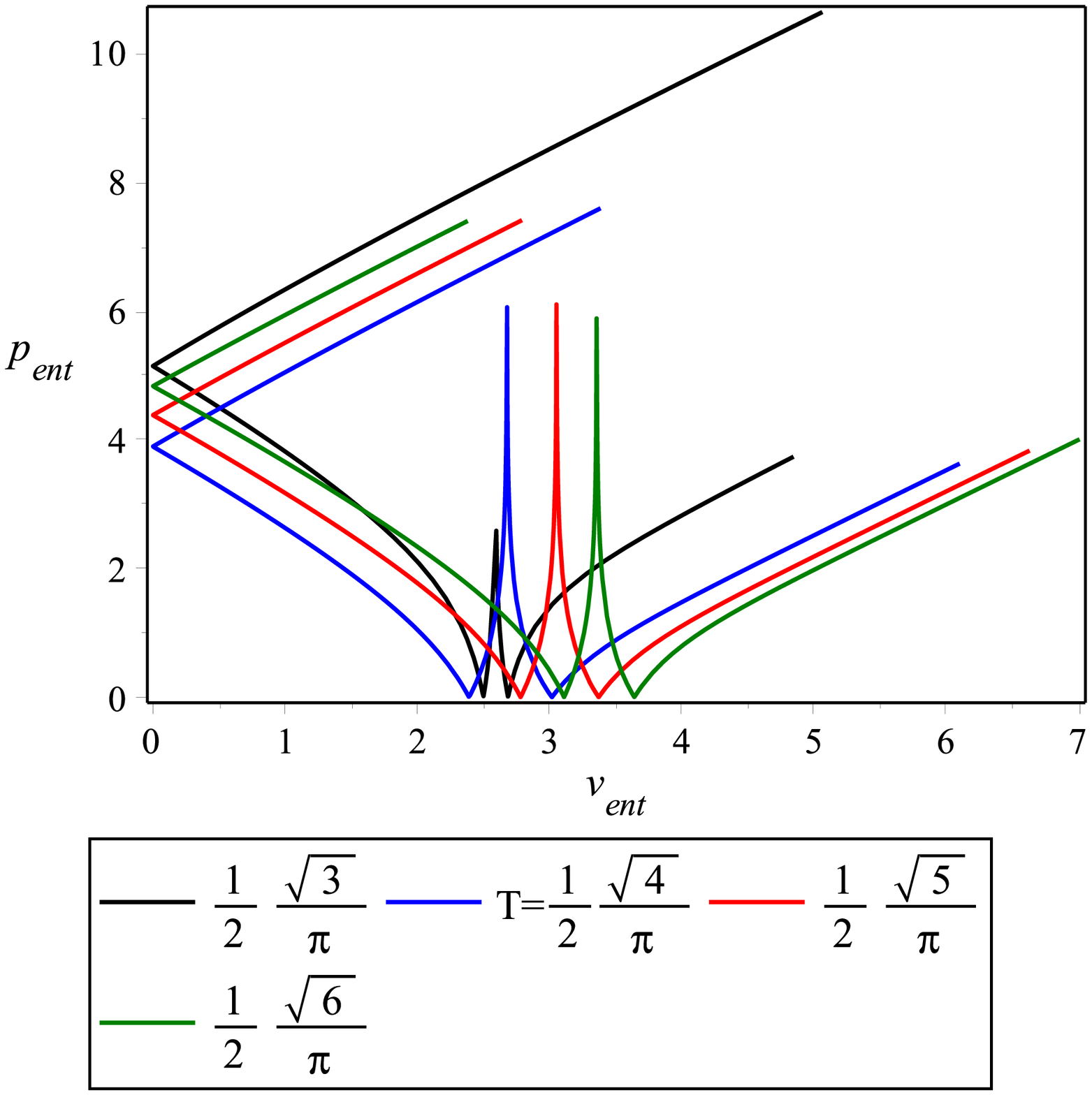}
  \caption{Graph of $p_{ent}v_{ent}$, given by Eqs. (\ref{pent}, \ref{Vcomplex}). }
\label{fig:complexity}
  \end{figure}

%%%%%%%%%%%%%%%%%%
In condensed matter physics, the fidelity susceptibility has been calculated for a  many-body 
quantum Hamiltonian  $H(\lambda)=H_0+\lambda H_I$,
where $\lambda$ is an external excitation 
parameter \cite{r6,r7, r8}. 
This
Hamiltonian can be diagonalized by an appropriate set of eigenstates 
$|n(\lambda)>$ and eigenvalues $E_m(\lambda)$,  
$ H(\lambda)|n(\lambda)>=E_n(\lambda)|n(\lambda)>.
$
These   eigenstates are usually taken as orthonormal  basis for the Hilbert space for CFT system. 
Now if two states   $\lambda,\lambda'=\lambda+\delta\lambda$
are  close to each other, then it is possible to define the distance between two states as   
$F(\lambda,\lambda+\delta\lambda)
=1- {\delta\lambda^2}\chi_F(\lambda)/2 +\mathcal{O}(\delta\lambda^4)
$, 
where the fidelity susceptibility of the system is  $\chi_F(\lambda)$  \cite{r6,r7, r8}.
This quantity $\chi_F(\lambda)$  can be  holographically calculated, and it is the equal to 
the informational complexity, when the volume is taken as the maximum volume in AdS, i.e.,  $V = V (\Sigma_{max})$ \cite{MIyaji:2015mia}. 
Now we can use this expression for any deformation of AdS, and here we will use it for SAdS$_4$. 
 So,  we can write the  volume term in this expression as 
\begin{eqnarray}
V(\Sigma_{max})=4\pi
\int^{r_{\infty}}_{r_{+}}
\frac{r^2dr}{\sqrt{f}}.
\label{fidelity}
\end{eqnarray}
It may be noted that if   
 we expand this expression in series, the zeroth order term   
 is divergent even for pure AdS. 
So, we can   define a fidelity volume $v_{fid}$, by subtracting   a volume term for pure AdS, $V(\Sigma_{max})_{AdS}$, from a  volume term for    AdS black holes,
$V(\Sigma_{max})_{SAdS}$, 
\begin{eqnarray}
 v_{fid} = V(\Sigma_{max})_{SAdS} - V(\Sigma_{max})_{AdS}\label{Vfid}. 
\end{eqnarray}
  To compare fidelity susceptibility with thermodynamics, we can use this 
fidelity volume $v_{fid}$. 
It may be noted that  in the extended phase space \cite{ext1, ext2}, a thermodynamic volume was defined as 
the quantity thermodynamically conjugate to the thermodynamic pressure (which was obtained using the cosmological constant). 
Here we will use the same argument to obtain the pressure conjugate to the  fidelity volume.  
So,  we can  define a new quantity,  which we call as the fidelity pressure. This quantity 
will be defined to be thermodynamically conjugate to the fidelity volume,  
\begin{eqnarray}
p_{fid}=-\frac{\frac{\partial M}{\partial r_{+}}}{\frac{\partial v_{fid}}{\partial r_{+}}}
\label{Pfid}
\end{eqnarray}
We numerically plotted it in  Fig.  (\ref{fig:fid}). This graph shows that there is a criticality in the $p_{fid}v_{fid}$.
This has the same form as the graph for the thermodynamic pressure and volume. 
This graph has been plotted using numerical integration to obtain  $v_{fid},p_{fid}$. 
To obtain numerical solutions the    cutoff parameter $r_{\infty}$   we will set it equal to 
$ {1}/{\delta},\delta\ll0.05$ in the numerical computations.  We numerically constructed an EoS for fidelity concept.
To compare the results with thermodynamic description  given in Fig. (\ref{fig:thermo}), we plotted the   fidelity pressure and fidelity 
volume  based on this EoS in the same isothermal regimes. 
It is observed that the fidelity pressure again increases with the fidelity volume till it reaching a maximum value. After reaching this
maximum value, it reduces with further increase in the volume. 
Thus, the thermodynamics of black holes and the fidelity susceptibility seem to represent the same physical process. However, the fidelity
susceptibility is well defined in terms of a boundary conformal 
field theory \cite{r6,r7, r8, MIyaji:2015mia}, and this would in principle imply that the thermodynamics of black hole  would be well defined 
in terms of boundary field theory. In fact, the fidelity susceptibility represents the 
difficulty to extract information from a process, so it is more  important to understand the difficulty to extract information during the evaporation
of a black hole, rather than the  loss of information during the  evaporation of a black hole. 
The fidelity volume measures this difficulty to extract information during the evaporation of a black hole. The fidelity 
pressure would be the quantity conjugate to this quantity, and would measure the flow of this quantity with the change in the mass of the black hole,
during its evaporation. 
Thus, the fidelity volume and fidelity pressure can be important quantity  which could be used to analyze such a process. It may be noted that
recent studied on black hole information have suggested that 
even though the information 
  may not be  actually lost in a black hole,  it would be effectively lost, as it would be impossible to obtain it back from Hawking radition \cite{hawk, chao, moffat}.  
  This again seems to indicate the information paradox in a black hole should be represented 
  by fidelity volume and fidelity pressure, and it is more   important to understand the difficulty to recover the information which could be expressed 
  in terms of these quantities.

%%%%%%%%%%%%%%%%%%%%%%%%%%%%%%%%%%
As alternative definition for the informational complexity of the boundary theory have been made using a different 
definitions for the volume in the bulk AdS, we will also use this  definition to compare it to thermodynamical volume. 
Thus, we will also use the   volume enclosed by the same   minimal surface which was used to calculate the holographic 
entanglement entropy,  $V = V(\gamma)$, and compare this to the   thermodynamic volume \cite{Alishahiha:2015rta}. 
Now for  $M\neq 0$,  the area integral for metric (\ref{sch_metric}) is defined as  
\begin{eqnarray}
\mathcal{A}(\gamma_A) =2\pi\int_{0}^{\theta_0}r \sin\theta\sqrt{r^2+\frac{r'^2}{f}}d\theta
\label{generalA}
\end{eqnarray}
here $\cos\theta_0=\frac{\rho_0}{\sqrt{1+\rho_0^2}},\rho_0\sim l$
The    Euler-Lagrange  equation  for $r=r(\theta)$ is given by, 
\begin{eqnarray}
&&r''=r'^2\Big(\frac{f' }{2f}-\frac{r'\cot\theta }{fr^2}+\frac{3}{r}\Big)-r'\cot\theta+2r f\label{EL}
\end{eqnarray}
 here prime denotes derivative with respect to the $\theta$. 
So, the  volume integral can now be written as 
\begin{eqnarray}
V(\gamma)=2\pi\int_{0}^{\theta_0}d\theta\sin\theta
\Big(\int^{r(\theta)}_{r_{+}}
\frac{r^2dr}{\sqrt{f}}
\Big)
\label{generalV}
\end{eqnarray}
To analyse the relation between the 
  holographic complexity and thermodynamics, we define  a  volume  as the 
  entanglement volume $v_{ent}$ and relate it to $V(\gamma)$. In fact, just as the fidelity volume, we define 
  this  entanglement volume $v_{ent}$  by subtracting this volume for pure AdS, $V(\gamma)_{AdS}$, from this volume for 
  AdS black hole, $V(\gamma)_{SAdS}$, 
  \begin{eqnarray}
   v_{ent} = V(\gamma)_{SAdS} - V(\gamma)_{AdS}\label{Vcomplex}. 
  \end{eqnarray}
  It may be noted that from the argument used in defining extended phase space \cite{ext1, ext2}, 
  fidelity pressure was defined to be thermodynamically conjugate to fidelity volume. 
  So, using the same argument,  
  we can define  the   entangled pressure $p_{ent}$ as a new quantity thermodynamically conjugate to the entanglement volume, 
\begin{eqnarray}
p_{ent}=-\frac{\frac{\partial M}{\partial r_{+}}}{\frac{\partial v_{ent}}{\partial r_{+}}}
\label{pent}
\end{eqnarray}
 Now we     numerically plot  $v_{ent}-p_{ent}$ for different values of temperature  in Fig.  
 (\ref{fig:complexity}).  It may be noted that the entanglement pressure can get negative, and so we use the absolute value of the pressure, 
 $|p_{ent}|$ in such a plot. 
To obtain the  numerical solution for holographic complexity, 
  we use the  the initial conditions $r(\theta_0)=\rho_0, $ and $ r'(0)=0$. 
We solve the  Euler-Lagrange  equation to  find $r(\theta)$, and 
obtain the holographic complexity.  

 Thus, we plot the     $v_{ent}-p_{ent}$, using numerical solutions for complexity pressure given by Eq.  (\ref{pent}) and its  conjugate volume . 
 It may be noted that  at peaks, we observed that $\frac{dp_{ent}}{dp_{ent}}\to \infty$. 
It may be noted that this graph diverges at points, and this behavior is expected, 
as we are using the same minimal surface which was used to calculate the holographic entanglement entropy, 
and such divergences have been observed to occur in holographic entanglement entropy \cite{hh12, hh15, hh14, hh17, hh18}. 
As we are using the same minimal surface, we would expect similar behavior for holographic complexity.  It would be interesting to find an explicit 
relation between the holographic complexity 
and entanglement entropy, as both these quantities are defined using the same minimal surface. Such a relation could be used to define the holographic 
complexity of a boundary theory. 
It is expected that it would measure that the entanglement volume could be used to analyze the   difficulty to extract information during a phase transition, 
and the entanglement pressure would 
indicate a holographic  flow in such a quantity, when the geometry describing such a quantities changes holographically.  It may be noted that this quantity 
does not resemble the behaviour of 
the thermodynamic volume and pressure. Thus, it would be more interesting to analyze the phase transition of a boundary theory using this quantity,
after defining its boundary dual rather than analyzing 
the black hole information paradox. However, as fidelity susceptibility does resemble  the behavior of  thermodynamic 
volume and pressure, the fidelity susceptibility would be the quantity to use for studying the black hole information paradox.

Thus, we have plotted various quantities which are represented by different definitions of the volume in AdS, 
and the conjugate  to these definitions of volume in AdS. For  each of these  cases,  we plotted the 
$PV$ graph for the same deformed AdS solution. Now we can compare the behavior of these different quantities using the 
 graphs  Figs.  (\ref{fig:fid},\ref{fig:complexity}) and (\ref{fig:thermo}). It was observed that behavior of  the  $PV$ graph for holographic complexity 
 was totally different from the $PV$ graphs for the thermodynamic volume   and fidelity susceptibility.  However, it  
  was also observed that the $PV$ graph  obtained from fidelity susceptibility
  and thermodynamic volume had almost the same behavior.    So, it was 
  conclude that to the  thermodynamical volume  in extended phase  and 
 fidelity susceptibility  represent the same physical quantity. The 
fidelity susceptibility can identified with informational complexity of the boundary theory, which is obtained geometrically using maximum volume in AdS.
So, in this paper, informational complexity was related to the thermodynamic volume of a theory using the the maximum volume in AdS spacetime. 
  Thus, the results of this paper established
a connection between geometry, thermodynamics, and information theory. It would be interesting to investigate this 
relation further, and analyse it for other deformations of the AdS spacetime.

 \end{document}